\documentstyle[preprint,aps,psfig,epsfig,floats]{revtex}
\begin{document}
\def\be{\begin{equation}}
\def\ee{\end{equation}}
\def\bc{\begin{center}}
\def\ec{\end{center}}
\def\bea{\begin{eqnarray}}
\def\eea{\end{eqnarray}}
\draft
\title{Competition and multiscaling in evolving networks}
\author{Ginestra Bianconi$^{*}$ and Albert-L\'aszl\'o Barab\'asi$^{*,\ \dag}$ }
\address{$^{*}$Department of Physics, University of Notre Dame, Notre Dame,
IN 46556, USA \\ $^{\dag}$ Institute for Advanced Studies,
Collegium Budapest,
 Szenth\'aroms\'ag utca 2,\\ H-1014 Budapest,
Hungary } 
\maketitle
\thispagestyle{empty}
\begin{abstract}
     The rate at which nodes in a network
increase their connectivity depends on their fitness to compete
for links. For example, in social networks some individuals
acquire more social links than others, or on the www some webpages
attract considerably more links than others. We find that this
competition for links translates into multiscaling, i.e. a fitness
dependent dynamic exponent, allowing fitter nodes to overcome the
more connected but less fit ones. Uncovering this
fitter-gets-richer phenomena can help us understand in
quantitative terms the evolution of many competitive systems in
nature and society.
\end{abstract}
\pacs{PACS numbers: 5.65+b, 89.75-k, 89.75Fb, 89.75Hc.}

The complexity of many systems can be attributed to the interwoven
web in which their constituents interact with each other. For
example, the society is organized in a social web, whose nodes are
individuals and  links represent various social interactions, or
the www forms a complex web whose nodes are documents and links
are URLs. While for a long time these networks have been modeled
as completely random $\cite{ER,Bollobas}$, recently there is
increasing evidence that they in fact have a number of generic
non-random characteristics, obeying  various scaling laws or
displaying short length-scale clustering
$\cite{Nature,Newrev,Pil,Sci,HubermannNature,BAScience,Faloutsos,Redner,RekaAlbpreprint,Porto,Amaralcondmat,WattsStrogatz,BarthelemyAmaral,Maritan}$.

A generic property of these complex systems is that they
constantly evolve in time. This  implies that the underlying
networks are not static, but continuously change through the
addition and/or removal of new nodes and links. Such evolving
networks  characterize the society, thanks to the birth and death
of nodes and their constant acquisition of social links;
 or characterize the www, where the number of nodes
 increases exponentially and the links connecting
 them are constantly modified $\cite{Nature,Sci,HubermannNature}$.
 Consequently, in addressing  these
complex systems we  have to uncover the dynamical forces that act
at the level of individual nodes, whose cumulative effect
determines the system's large-scale topology. A first step in this
direction was the introduction of the scale-free model
$\cite{BAScience}$, that incorporates the fact that
network evolution is driven by at least two coexisting mechanisms:
(1) growth, implying that networks continuously expand by the
addition of new nodes that attach to the nodes already present in
the system; (2) preferential attachment, mimicking the fact that a
new node links  with higher probability to the nodes that already
have a large number of links. With these two ingredients the
scale-free model predicts the emergence of a power-law
connectivity distribution, observed in many systems
$\cite{Nature,BAScience,Faloutsos,Redner}$, ranging from the
Internet to citation networks.
 Furthermore, extensions of this model, including rewiring
$\cite{RekaAlbpreprint}$ or aging $\cite{Porto,Amaralcondmat}$
have been able to account for more realistic aspects of the
network evolution, such as the existence of various scaling
exponents or cutoffs in the connectivity distribution.

Despite its success in predicting the large-scale topology of real
networks, the scale-free model neglects an important aspect of
competitive systems:  not all nodes are equally successful in
acquiring links. The model predicts that all nodes increase their
connectivity in time as $k_i(t) = (t/t_i)^{\beta}$, where
$\beta=1/2$ and $t_i$ is the time at which node $i$ has been added
into the system. Consequently, the oldest nodes will have the
highest number of links, since they had the longest timeframe to
acquire them. Since new nodes attach preferentially to more
connected nodes, highly connected nodes will continue to acquire
further links at a higher rate at the expense of the smaller
nodes, a rich-gets-richer phenomena which is responsible for the
power-law tail of the connectivity distribution.
  Furthermore, if two nodes arrive at the same time,
  apart from some statistical fluctuations,
  at any time they will approximately have the same number of links.

 On the other hand numerous examples convincingly indicate
 that in real systems
a node's connectivity and growth rate does not depends on its age
alone. For example, in social systems not everybody makes friends
at the same rate: some individuals are better in turning a random
meeting into a lasting social link than others. On the www some
documents through a combination of good content and marketing
acquire a large number of links in a very short time, easily
overtaking websites that have been around for much longer time
$\cite{Hubermanncond-mat}$. Also, in Hollywood some actors in  a
very short timeframe build a movie portfolio and a collection of
links that easily surpasses many actors in business for much
longer time. Finally, some research papers in a short timeframe
acquire a very large number of citations, much in excess of the
majority of their contemporary or even older publications. In all
these examples we see a similar pattern: some nodes acquire links
at a rate much higher than other nodes in the system. We tend to
associate these differences with some intrinsic quality of the
nodes, such as the social skills of an individual, the content of
a web page, the talent of an actor or the content of a scientific
article. We will call this the node's fitness, describing it's
ability to compete for links at the expense of other nodes. While
such competition for links in real systems is well documented, it
has not been incorporated  in the current network models. In this
paper we take a first step in this direction by proposing a simple
model that allows us to investigate this competitive aspect of
real networks in quantitative terms. Assuming that the existence
of a fitness modifies the preferential attachment to compete for
links, we find that different fitness translates into multiscaling
in the dynamical evolution: while the connectivity of individual
nodes will still follow a power-law in time, i.e. $k_i(t)\sim
t^{\beta_i}$, the dynamical exponent, $\beta_i$, will depend on
the fitness of the node. We develop the continuum model for this
competitive evolving network, allowing us to calculate
analytically $\beta$ and derive a general expression for the
connectivity distribution. We find that the analytical predictions
are in excellent agreement with the results obtained from
numerical simulations.

{\it The fitness model --} The examples discussed above indicate
that nodes have different ability  (fitness) to compete for links.
To account for these differences  we introduce a fitness
parameter, $\eta_i$, that we assign to each node, and assume that
it is unchanged in time (i.e. $\eta_i$ represents a quenched
noise) $\cite{nota1}$. Starting with a small number of nodes, at
every timestep we add a new node $i$ with fitness $\eta_i$, where
$\eta$ is chosen from the distribution $\rho(\eta)$. Each new node
$i$ has $m$ links that are connected to the nodes already present
in the system. We assume that the probability $\Pi_i$ that a new
node will connect to a node $i$ already present in the network
depends on the connectivity $k_i$ and
 on the fitness $\eta_i$ of that node, such that
\be
\Pi_i=\frac{\eta_i k_i}{\sum_j \eta_j k_j}. \label{p} \ee This
generalized preferential attachment $\cite{BAScience}$
incorporates in the simplest possible way that fitness and
connectivity  jointly determine the rate at which new links are
added to a given node, i.e. even a relatively young node with a
few links can acquire links at a high rate if it has a large
fitness parameter.
 To address
the scaling properties of this model we  first develop a continuum
theory, allowing us to predict the connectivity distribution
$\cite{BAScience,RekaAlbpreprint,Porto}$. A node $i$
will increase its connectivity $k_i$ at a rate that is
proportional to the probability ($\ref{p}$) that a new node will
attach to it, giving
\be
\frac{\partial k_i}{\partial t}= m \frac{\eta_{i} k_{i}}{\sum_{j}
k_{j}\eta_{j}}. \label{din1} \ee The factor $m$ accounts for the
fact that each new node adds $m$ links to the system.
  If $\rho(\eta)=
\delta(\eta-1)$, i.e. all fitness are equal, ($\ref{din1}$)
reduces to
 the scale-free model, which predicts that $k_i(t) \sim t^{1/2}$
 $\cite{BAScience}$.
In order to solve ($\ref{din1}$) we assume that similarly to the
scale-free model the time evolution of $k_i$ follows a power-law,
but there is multiscaling in the system, i.e. the dynamic exponent
depends on the fitness $\eta_{i}$,
\be
k_{\eta_i}(t,t_0)=m \left( \frac{t}{t_0} \right)^{\beta(\eta_i)},
\label{power} \ee where $t_0$ is the time at which  the node $i$
was born. The dynamic exponent
 $\beta(\eta)$ is bounded, i.e. $0<\beta(\eta)<1$ because
 a node always increases the number of links in time ($\beta(\eta)>0$)
and $k_i(t)$ cannot increases faster than $t$ ($\beta(\eta)<1$).
 We first calculate the mean of the sum
$\sum_j \eta_j k_j$ over all  possible realization of
 the quenched noise ${\{\eta\}}$.
 Since each node is born at a different time $t_0$,
the sum over $j$ can be  written as an integral over $t_0$
 \bea <\sum_j \eta_j k_j>&=&\int d\eta
\rho(\eta)\ \eta
 \int_1^t dt_0\  k_{\eta} (t,t_0)\nonumber\\
&=&\int d\eta\  \eta \rho(\eta) m \
\frac{(t-t^{\beta(\eta)})}{1-\beta(\eta)}. \eea Since
$\beta(\eta)<1$, in the $t \rightarrow \infty$ limit
$t^{\beta(\eta)}$ can be neglected compared to $t$, thus we obtain
\be
 <\sum_j \eta_j k_j>\stackrel{t\rightarrow\infty}{=}  C m t
 (1+O(t^{-\epsilon}),
\label{5} \ee where \bea \epsilon=(1-\max_{\eta}\beta(\eta))>0,
\nonumber \\
 C=\int {d\eta \rho(\eta) \frac{
\eta}{1-\beta(\eta)}} . \label{C1} \eea Using ($\ref{5}$), and the
notation $k_{\eta}=k_{\eta_i}(t,t_0)$ the dynamic equation
($\ref{din1}$) can be written as
\be
\frac{\partial k_{\eta}}{\partial t}= \frac{\eta k_{\eta}}{C t},
\label{dinc} \ee
which has a solution of form ($\ref{power}$),
given that
\be
\beta(\eta)=\frac{\eta}{C}, \label{7} \ee thereby confirming the
self-consistent nature of the assumption ($\ref{power}$). To
complete the calculation we need to determine
  $C$ from ($\ref{C1}$) after substituting
$\beta(\eta)$ with $\eta/C$,
\be
1= \int_0^{\eta_{max}} d\eta \rho(\eta)
\frac{1}{\frac{C}{\eta}-1}, \label{selfc} \ee where $\eta_{max}$
is the maximum possible fitness in the system $\cite{massc}$.
Apparently ($\ref{selfc}$) is a singular integral. However, since
$\beta(\eta)=\eta/C<1$ for every value of $\eta$, we have
$C>\eta_{max}$, thus the integration limit never reaches the
singularity. Note also that, since $\sum_j \eta_j k_j\le
\eta_{max} \sum_j k_j =2 m t \eta_{max}$, we have, using
($\ref{5}$), that $C\le 2 \eta_{max}$.

Finally, we can  calculate the connectivity distribution $P(k)$,
which gives the probability that  a node has $k$ links. If there
is a single dynamic exponent $\beta$, the connectivity
distribution follows the power-law $P(k) \sim k^{\gamma}$, where
the connectivity exponent is given by $\gamma=1/\beta +1$.
However, in
 this model we have  a spectrum of dynamic exponents
$\beta(\eta)$, thus $P(k)$ is given by a weighted sum over
different power-laws. To find $P(k)$ we need to calculate the
cumulative probability that for a certain node  $k_{\eta}(t)>k$,
\bea P(k_{\eta}(t)>k) &=& P\left(t_0<t\left(
\frac{m}{k}\right)^{C/\eta}\right)\nonumber \\ &=& t \ {\left(
\frac{m}{k}\right)}^{\frac{C}{\eta}}. \eea Thus the connectivity
distribution, i.e. the probability that a node has  $k$ links, is
given by the integral\bea P(k)&=&\int_0^{\eta_{max}} d\eta
\frac{\partial P(k_{\eta}(t)>k)}{\partial t}=\nonumber \\ &\propto
& \int d\eta \rho(\eta) \frac{C}{\eta}
\left(\frac{m}{k}\right)^{\frac{C}{\eta}+1}. \label{pk} \eea

{\it Scale-free model--} Given the fitness distribution
$\rho(\eta)$, the continuum theory allows us to predict both the
dynamics, described by the dynamic exponent $\beta(\eta)$ (Eqs.
(\ref{7}) and ($\ref{selfc}$)), and the topology, characterized by
the connectivity distribution $P(k)$ (Eq. (\ref{pk})). To
demonstrate the validity of our predictions, in the following we
calculate these quantities for two different $\rho(\eta)$
functions. As a first application let us consider the simplest
case, corresponding to the scale-free model, when all fitnesses
are equal. Thus we have $\rho(\eta)=\delta(\eta-1)$, which,
inserted in ($\ref{selfc}$), gives  $C=2$, which represents  the
largest possible value of $C$. Using ($\ref{7}$) we obtain
$\beta=1/2$ and from ($\ref{pk}$) we get $P(k)\propto k^{-3}$, the
known scaling of the scale-free model. Thus the scale-free model
represents an extreme case of the fitness model considered here,
the connectivity exponent taking up the largest possible value of
$\gamma$.

{\it Uniform fitness distribution--} The behavior of the system is
far more interesting, however, when nodes with different fitness
compete for links. The simplest such case, which already offers
nontrivial multiscaling, is obtained when $\rho(\eta)$ is chosen
uniformly from the interval $[0,1]$. The constant $C$ can be
determined again from ($\ref{selfc}$), which gives
\be
\exp(-2/C)=1-1/C, \label{14} \ee whose solution is $C^*=1.255$.
Thus, according to ($\ref{7}$), each node will have a different
dynamic exponent, given by $\beta(\eta)\sim \frac{\eta}{C^*}$.
Using ($\ref{pk}$)  we obtain
\be
P(k) \propto  \int_0^1 d\eta  \frac{C^*}{\eta}
\frac{1}{k^{1+C^*/\eta}} \sim  \frac{k^{-(1+C^*)}}{\log(k)},
\label{pl} \ee i.e. the connectivity distribution follows a
generalized power-law, with an inverse logarithmic correction.

\begin{figure}
\centerline{\epsfxsize=3.5in \epsfbox{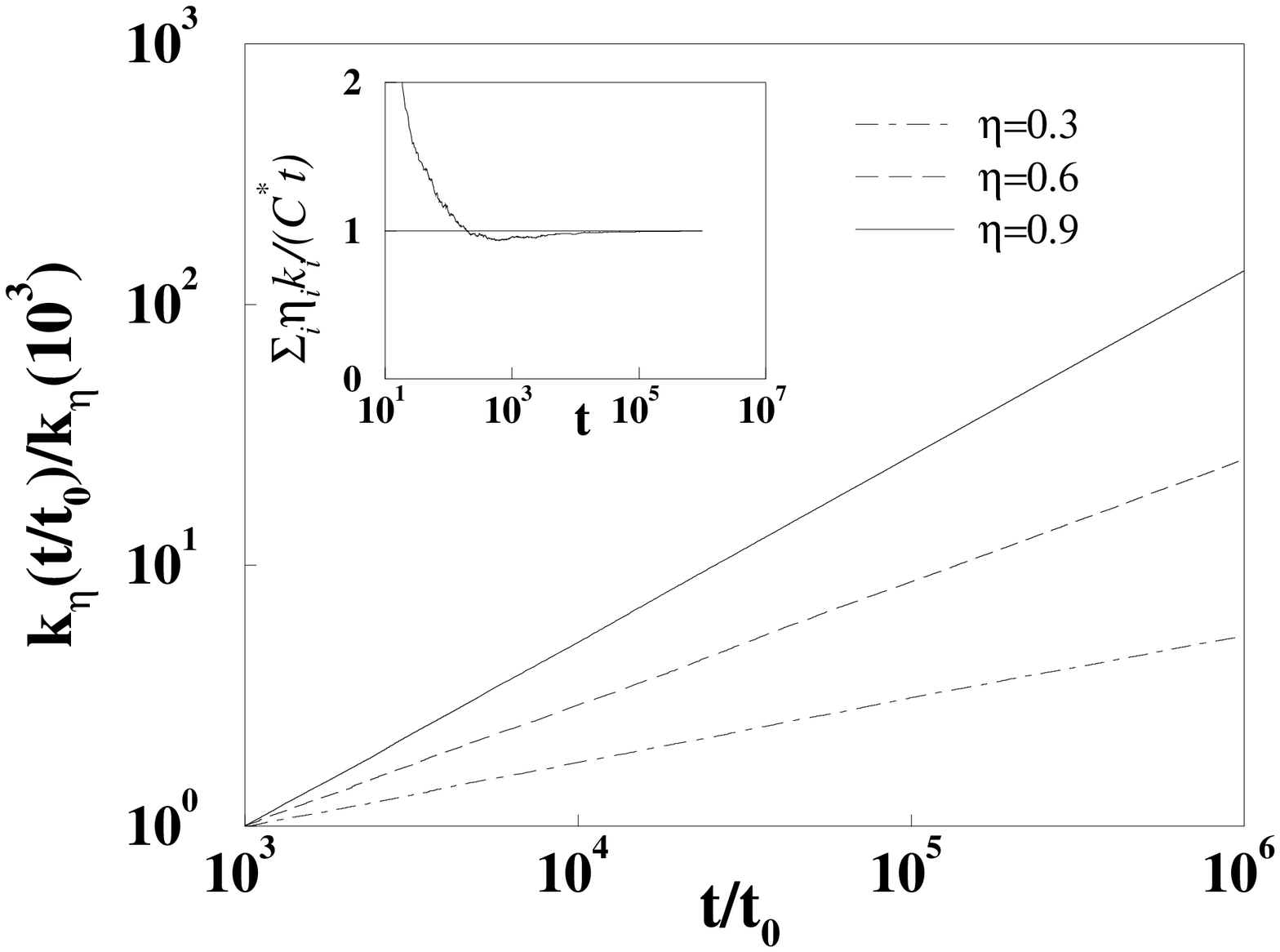}} 
\caption{Time
dependence of the connectivity, $k_{\eta}(t)$, for nodes with
fitness $ \eta =0.3,\ 0.6$ and $0.9$. Note that $k_{\eta}(t)$
follows a power-law in each case and  the dynamic exponent
$\beta(\eta)$, given by the slope of $k(t)$, increase with $\eta$.
While in the simulation the fitness of the nodes have been drawn
uniformly, between $[0,1]$, in the figure we show only the
connectivity of three nodes with selected fitness. In the
simulation we used $m=2$ and the shown curves represent averages
over 20 runs. {\it Inset}: Asymptotic convergence of
$(\sum_{i=1}^t \eta_i k_i)/t$ to the analytically predicted limit
$C^*=1.255$, shown as an horizontal line (see Eq. ($\ref{14}$)).}
\label{1fig}
\end{figure}

\begin{figure}
\centerline{\epsfxsize=3.5in \epsfbox{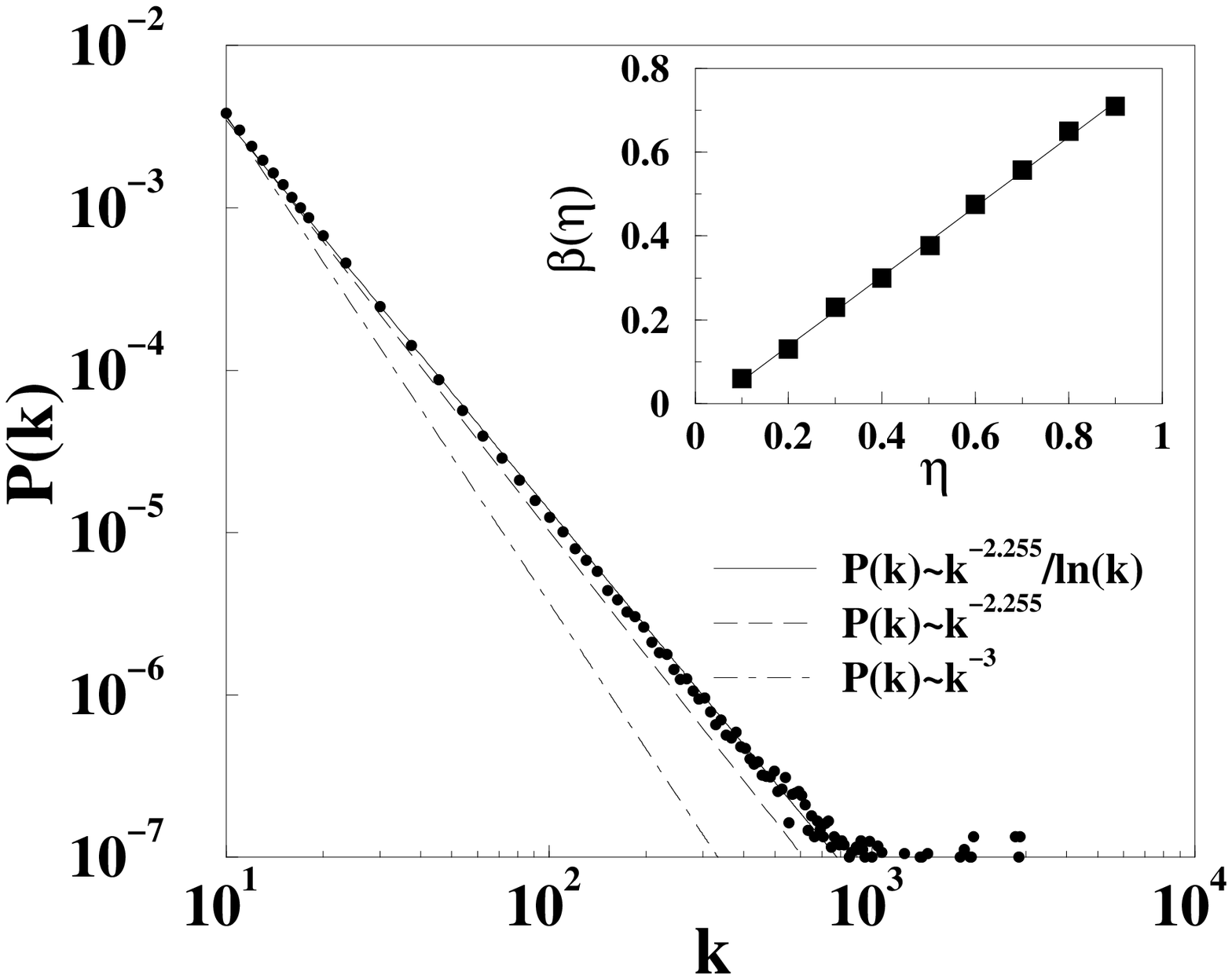}} 
\caption{
Connectivity distribution in the fitness model, obtained for a
network with $m=2$ and $N=10^6$ nodes. The upper solid line that
goes along the circles provided by the numerical simulations
corresponds to the theoretical prediction (\ref{pl}), with
$\gamma=2.25 $. The dashed line corresponds to a simple fit
$P(k)\sim k^{-2.255}$ without the logarithmic correction, while
the long-dashed curve correspond to $P(k)\sim k^{-3}$ , as
predicted by the scale-free model,
in which all fitness are equal. {\it Inset}: The dependence of the
dynamic exponent $\beta(\eta)$ on the fitness parameter $\eta$ in
the case of a uniform $\rho(\eta)$ distribution. The squares were
obtained from the numerical simulations while  the solid line
corresponds to the analytical prediction 
$\beta (\eta )= \eta /1.255$.}
\label{2fig}
\end{figure}

 To check the predictions of the continuum theory we
performed
 numerical simulations of the discrete fitness model, choosing
 fitness with equal probability from the interval $[0,1]$.
Most important is to test the validity of the ansatz
($\ref{power}$), for which we recorded the time evolution of nodes
with different fitness $\eta$. As Fig.~$\ref{1fig}$ shows, we find
that
 $k_i(t)$ follows a power-law for all
$\eta$, and  the scaling exponent, $\beta(\eta)$, depends on
$\eta$, being larger for nodes with larger fitness. Eq.
($\ref{C1}$) predicts that the sum $< \sum_i \eta_i k_i>/m t
\rightarrow C^*$ in the $t \rightarrow \infty$ limit, where $C^*$
is given by ($\ref{14}$) as $C^*=1.255$. Indeed  as the inset in
Fig.~$\ref{1fig}$ shows, the discrete network model indicates that
this sum converges to the analytically predicted value.
Figure~$\ref{1fig}$ allows us to determine numerically the
exponent $\beta(\eta)$, and compare it to the prediction
($\ref{7}$). As the inset in Fig.~$\ref{2fig}$ indicates, we
obtain excellent agreement between the numerically determined
exponents and the prediction of the continuum theory. Finally, in
Fig.~$\ref{2fig}$ we show the agreement between the prediction
($\ref{pl}$) and the numerical results for the connectivity
distribution $P(k)$.

An interesting feature of the numerically determined connectivity
distribution (Fig.~$\ref{2fig}$) is the appearance of a few nodes
that have higher number of links than predicted by the
connectivity distribution. Such highly connected hubs, appearing
as a horizontal line with large $k$ on the log-log plot, are
present in many systems, including the www $\cite{Nature}$ or the
metabolic network of a cell $\cite{cell}$, clearly visible if we
do not use logarithmic binning. This indicates, that the
appearance of a few "super hubs", i.e. nodes that have connections
in excess to that predicted by a power-law, is a generic feature
of competitive systems.

{\it Discussion --} The above results offer interesting insights
into the evolution of  nodes in a
  competitive environment. The model studied by us reflects the basic
  properties of many real systems in which the nodes
  compete for links
  with other nodes, thus a node can acquire links only
at the expense of the other nodes.
 The competitive nature of the model
is guaranteed by the fact that the new node has  only a fixed
number of links, $m$, which, as the system grows, are distributed
between an increasing number of nodes. Thus nodes that are already
in the system have to compete with a linearly increasing number of
other nodes for a link.
    In the scale-free model, where each node
   has the same fitness, all nodes increase their connectivity following
   the same scaling exponent $\beta=1/2$.
 In contrast, we find that when we allow different fitness,
 multiscaling emerges and
  the dynamic exponent depends on the fitness parameter, $\eta$.
This allows nodes with a higher fitness to enter the system at a
later time and overcome nodes that have been in the system for a
much longer timeframe. Such nodes with a higher fitness correspond
to people with higher social skills; to websites with better
content or services; or to articles that by report some important
discovery. What is interesting, however, is that despite the
significant differences in their fitness, all nodes  will continue
to increase their connectivity following a power-law in time.
Thus, our results indicate that the fitter wins by following a
power-law time dependence with a higher exponent than its less fit
peers $\cite{nota2}$.

 Beyond the conceptual importance of these
results, the predictions of the model could be verified  on
networks for which the dynamic evolution of the nodes can be
extracted, such as the science citation index or the actor network
(for both of which the date at which a node is added to the system
is recorded). Such measurement could also offer an independent
determination of the fitness parameters and $\rho(\eta)$, which
would allow the simultaneous testing of ($\ref{power}$),
($\ref{7}$) and ($\ref{pk}$).

We wish to acknowledge useful discussion  with R. Albert, I.
Der\'enyi, H. Jeong, E. Szathm\'ary, T. Vicsek. This research was
partially supported by NSF Career Award DMR-9710998.


\newpage

\begin{thebibliography}{99}
\bibitem{nota1}
Note that in some real systems the fitness can change with time,
for example a research field can slowly close down or an actor can
suspend acting diminishing the ability of the corresponding nodes
to compete for links $\cite{Porto,Amaralcondmat}$.
\bibitem{massc}
Equation ($\ref{selfc}$) can be also derived  from the
normalization condition $2 k_0 t=\sum_{j\epsilon N(t)} k_j$, a
"mass conservation"  law, giving the total number of links in the
network at time $t$.
\bibitem{nota2}
In many systems, such as the www, nodes can acquire links through
rewiring or the appearance of new internal links
$\cite{RekaAlbpreprint}$. Such processes can be included in the
present model, enhancing further the connectivity of the fittest
nodes.
\end{thebibliography}
\end{document}